\documentstyle[12pt,epsfig,a4]{article} 
\newlength{\dinwidth} 
\newlength{\dinmargin} 
\def\lapproxeq{\lower .7ex\hbox{$\;\stackrel{\textstyle<}{\sim}\;$}} 
\def\gapproxeq{\lower .7ex\hbox{$\;\stackrel{\textstyle>}{\sim}\;$}} 
                                          
\def\be{\begin{equation}}                                    
\def\ee{\end{equation}}                                   
\def\bea{\begin{eqnarray}}                                    
\def\eea{\end{eqnarray}}                                                                                             
 
\def\qbar{\bar q} 
\def\ftil{\hat{\cal{F}}}

\def\Xprime{X^{\prime}} 
\def\Zprime{Z^{\prime}}

\def\xprim2bar{\overline{x}^{\prime\prime}}

\makeatletter                                                 
\def\fmslash{\@ifnextchar[{\fmsl@sh}{\fmsl@sh[0mu]}}  
\def\fmsl@sh[#1]#2{%
\mathchoice                                                 
{\@fmsl@sh\displaystyle{#1}{#2}}%
{\@fmsl@sh\textstyle{#1}{#2}}%
{\@fmsl@sh\scriptstyle{#1}{#2}}%
{\@fmsl@sh\scriptscriptstyle{#1}{#2}}}                                    
\def\@fmsl@sh#1#2#3{\m@th\ooalign{$\hfil#1\mkern#2/\hfil$\crcr$#1#3$}}                                                     
 
\makeatother                                                                                                          
\begin{document}      
\titlepage            
\begin{flushright}                                                 
DTP/98/48 \\                                                 
July 1998 \\                                                 
\end{flushright}

\begin{center}                                            
      
\vspace*{2cm}                                             
     
{\Large \bf Off-diagonal parton distributions and their evolution}

\vspace*{1cm}                                             
     
K.J.~Golec-Biernat\footnote{On leave from H.~Niewodniczanski Institute of 
 Nuclear    
Physics, ul.~Radzikowskiego 152, Krakow, Poland.} and A.D.~Martin

\vspace*{0.5cm}

Department of Physics, University of Durham, Durham, DH1 3LE, UK.         
                          
\end{center}

\vspace*{2cm}

\begin{abstract}                                          
        
We construct off-diagonal parton distributions defined on the interval $0 \le X \le 1$  
starting from the off-forward distributions defined by Ji. 
We emphasize the particular role played by the 
symmetry  relations in the \lq\lq ERBL-like" region.  
We find the evolution equations for the off-diagonal distributions which 
conserve these symmetries.  
We present numerical results  of the evolution, and  
verify that the analytic asymptotic forms of the parton distributions 
are reproduced.  We also compare the constructed off-diagonal distributions with the 
non-forward distributions defined by Radyushkin and   
comment on the singularity structure 
of the basic amplitude written in terms of the  off-diagonal distributions. 
        
\end{abstract}                                                                                                          
\newpage

\section{Introduction} 
                    
It is well known that the cross section of hard scattering processes (such as deep                                                
inelastic scattering, the production of large $p_T$ jets, etc.) can be written as the sum                                                
of parton distributions multiplied by the cross sections of hard subprocesses calculated                                                
at the parton level using perturbative QCD.  That is we can factor off the long distance                                                
(non-perturbative) effects into universal, process independent, parton distributions                                                
($f_i (X, \mu^2)$ with $i = q, \bar{q}, g$) specific to the incoming hadrons.  $X$ is    
the longitudinal fraction of the hadron's momentum that is carried by the parton and    
$\mu$ is a scale typical of the hard subprocess.  The parton distributions are given by    
the matrix elements $\langle P | \hat{O} | P \rangle$ where $\hat{O}$ is a twist-2    
quark or gluon operator, and $P$ represents the full set of quantum numbers of the    
hadron.  To be specific we will be concerned with a proton taking part in    
unpolarised reactions.  Thus $P$ will represent the 4-momentum of the proton.   
                                               
Calculating the parton distributions from first principles is one of the most challenging                                                
problems in non-perturbative QCD.  The most promising approach is lattice QCD, but                                                
much remains to be done.  On the other hand, from a practical viewpoint, the parton                                                
distributions of the proton are determined with good precision from global analyses of    
deep inelastic and related hard scattering data.  The distributions $f_i (X, \mu^2)$ are                                                
parametrized as a function of $X$ at some starting scale $\mu_0^2$ and then                                                
evolved using the DGLAP equations of perturbative QCD to higher $\mu^2$ values                                                
relevant to the data to be fitted.                                               
                                               
Recently \cite{JI}--\cite{GKM} 
there has been much interest in off-diagonal (also called off-forward by Ji    
\cite{JI} or non-forward by Radyushkin \cite{RAD}) distributions which are given by 
matrix elements 
$\langle P^\prime|\hat{O}| P \rangle$ in which the momentum of the outgoing proton 
is  
not the                                                
same as that of the incoming proton.  For example, the {\it amplitudes} for processes                                                
such as deeply virtual Compton scattering $(\gamma^* p \rightarrow \gamma p)$ or                                                
vector particle electroproduction $(\gamma^* p \rightarrow Zp$ or $J/\psi p$) depend                                                
on off-diagonal distributions.  Since $P \neq P^\prime$ the parton returning to the    
proton has a different momentum to the one which is outgoing, and so we need two    
momentum variables to specify the off-diagonal distributions.  The Ji and Radyushkin    
distributions, which are denoted by $H (x, \xi)$ and ${\cal{F}}_\zeta (X)$    
respectively, differ in their choice of the defining four vector.  Ji chooses the    
momentum fractions $x$ and $\xi$ with respect to the average of the incoming and    
outgoing proton momenta $\bar{P} = \frac{1}{2} (P + P^\prime)$, whereas    
Radyushkin defines $X$ and $\zeta$ with respect to the incoming proton momentum    
$P$.  The former has the important advantage that it is easier to impose the symmetry    
requirements, while the latter has the advantage that it is close to the definition used    
for the conventional (diagonal) distributions. Our aim is to clarify 
the relation between the two formulations. We find that 
they are not equivalent unless specific 
conditions are imposed on Radyushkin's non-forward distributions. We show this  
by a direct construction of distributions defined in the range $0 \le X \le 1$ which are 
equivalent to  Ji's off-forward distributions.  
   
Let us neglect, for the moment, the gluon distribution.   
The quark distribution $H_q(x, \xi)$, defined by Ji, 
covers the interval $-1 \le x \le 1$  and 
generates two distinct distributions    
which we denote\footnote{For the reasons given below we must use a notation
which distinguishes between the distributions $\ftil(X,\zeta)$
constructed from $H$ and the non-forward 
distributions ${\cal{F}}_{\zeta}(X)$ defined by Radyushkin.}  
by $\hat{\cal{F}}_q (X, \zeta)$ and $\hat{\cal{F}}_{\bar{q}}(X,\zeta)$  
with $0 \le X \le 1$.  Over the region $X>\zeta$ the two functions 
$\hat{\cal{F}}_q$ and $\hat{\cal{F}}_{\bar{q}}$ are independent.  
On the other 
hand in the region $X<\zeta$  
they are related to each other, with the consequence that 
the non-singlet and singlet combinations possess a symmetry 
about $X = \zeta/2$.  We obtain evolution equations for $\hat{\cal{F}}$ 
starting from the evolution equations for the off-forward distributions $H$. 
We find that they differ from the evolution equations for  
the non-forward distributions \cite{RAD2,GKM} 
by additional terms which are essential to preserve the symmetry 
properties in the ERBL-like region. We also found that the basic amplitude 
for deeply virtual Compton scattering (DVCS) has a different singularity structure 
to that given by the non-forward distributions ${\cal{F}}$. 
   
The outline of the paper is as follows.  To establish notation we quickly    
review in Section 2 the salient features of the conventional (diagonal) parton    
distributions $H (x)$ with support $-1 \le x \le 1$.  Section 3 reviews the extension of    
these ideas to the off-diagonal distributions $H (x, \xi)$ that were 
introduced by Ji \cite{JI}.   
In   Section 4 we transform the distributions $H (x, \xi)$ into distributions    
$\hat{\cal{F}} (X, \zeta)$ with $0 \le X \le 1$, and demonstrate that    
$\hat{\cal{F}}$ must satisfy symmetry relations for $X < \zeta$.  In Section 5 we    
give the evolution equations for the $\hat{\cal{F}} (X, \zeta)$ and present numerical    
solutions.  The complete form of the evolution equations  
is given in the Appendix. In Section 6 we discuss the relation between the 
distributions $\hat{\cal{F}} $ 
and the non-forward distributions ${\cal{F}}$ of Radyushkin.   
In the same spirit we discuss the differences  
in the singularity  structure of the DVCS amplitude. 
Finally Section 7 contains our conclusions.

\section{Conventional parton distributions} 
 
In order to introduce off-diagonal distributions it is most convenient to first recall the 
definition of the conventional (diagonal) parton distributions in terms of light-cone 
coordinates $(x^\pm = (x^0 \pm x^3)/\sqrt{2}, x^1, x^2)$ and in the light-cone gauge    
$(A^+ = 0)$ \cite{DIAG}.  For instance the quark distribution $H_q (x)$ is given in    
terms of the matrix element of a light-cone bilocal operator              
\be 
\label{eq:a2} 
H_q (x)  \; = \; \frac{1}{2} \int \frac{d y^-}{2 \pi} e^{-i x P^+ y^-}  
\: \langle P |   
\bar{\psi}_q \! ( 0, y^{-}/2, \mbox{\boldmath $0$} ) \: \textstyle{\frac{1}{2 
}} \gamma^+    
\: \psi_q \! \left ( 0, -y^{-}/2, \mbox{\boldmath $0$} \right ) | P \rangle .      
\ee 
Note that the matrix element is diagonal in the four momentum $P$ of the proton.  For   
simplicity we do not show either here, or throughout the paper,  
the renormalization scale dependence of $H_q$ 
and of the other parton distributions that we discuss.    
             
To see the parton content of the distribution $H_q$ we make a Fourier expansion of 
(the light-cone-plus or \lq good' component) $\psi_+$ of the quark field, in terms of         
the quark annihilation operator $b$ and the antiquark creation operator $d^\dagger$.          
Similarly $\bar{\psi}_+$ is expanded in terms of $b^\dagger$ and $d$, and then the         
integration over $y^-$ in (\ref{eq:a2}) is performed. It is found                
that $H_q$ is only non-vanishing in the interval $-1 \le x \le 1$ with the term 
$b^\dagger b$ contributing for $x > 0$ and $dd^\dagger$ contributing for $x < 0$ 
\cite{JAFFE}  
\bea  
\label{eq:b2}  
H_q (x) & = & \frac{1}{2P^+} \: \int \:  
\frac{d^2 k_T}{2x (2 \pi)^3} \: \sum_\lambda   
\left [ \langle P | b_\lambda^\dagger \! (x P^+, \mbox{\boldmath $k$}_T) \:   
b_\lambda \! (x P^+, \mbox{\boldmath $k$}_T ) | P \rangle \:  
\theta (x) \right. \nonumber \\  
& & \\  
& & - \! \left . \langle P | d_\lambda^\dagger \!  
( -x P^+, \mbox{\boldmath $k$}_T ) \:    
d_\lambda \! ( -x P^+, \mbox{\boldmath $k$}_T ) | P \rangle \: \theta (-x)\;,   
\right ] \nonumber  
\eea  
where $\lambda$ is the helicity of the quarks.  The $b^\dagger b$ term corresponds to   
the emission of a quark (carrying a fraction $x$ of the proton's momentum) and its   
subsequent reabsorption within the proton.  Similarly the $d^\dagger d$ contribution   
describes the emission and subsequent reabsorption of an antiquark.  The two   
possibilities are sketched in Fig.~1.  Thus the single distribution $H_q$ with support   
in the interval $-1 \le x \le 1$ embodies both the familiar $q$ and $\bar{q}$   
distributions, defined on the interval $0 \le x \le 1$, which thus are identified with the         
two terms accompanying the theta functions in (\ref{eq:b2}) in the following way        
\be             
\label{eq:a3}             
H_q (x) \; = \: \left \{ \begin{array}{lcl}      
q (x) & {\rm for} & x > 0 \\     
& & \\            
- \bar{q} (-x) & {\rm for} & x < 0. \end{array} \right .   
\ee                         
We may form the valence and singlet quark distributions in terms of $H_q$ 
\bea         
\label{eq:a4}         
q (x) \; - \; \bar{q} (x) & = & H_q (x) \; + \; H_q (-x) \; \equiv \; H 
_q^V (x)          
\nonumber \\         
& & \\         
\sum_q \: \left [ q (x) \; + \; \bar{q} (x) \right ] & = & \sum_q \left 
 [H_q (x) \: -    
\: H_q (-x) \right ] \; \equiv \; H^S (x) \;, \nonumber            
\eea         
where the sum is over the quark flavours.  Clearly over  
the full interval  $-1 \le x \le 1$    
the valence and singlet quark distributions satisfy the symmetry relations        
\bea        
\label{eq:a5}        
H_q^V (x) & = & H_q^V (-x) \nonumber \\                 
& & \\        
H^S (x) & = & - H^S (-x)\;. \nonumber        
\eea        
        
In a similar way we may introduce $H_g (x) \equiv xg (x)$ where $g (x)$ is the    
familiar gluon distribution.   In the light-cone gauge             
\be        
\label{eq:a6}        
H_g (x) \; = \; \frac{1}{P^+} \: \int \: \frac{dy^-}{2 \pi} \: e^{- i x 
 P^+ y^-} \: \langle  
P | F^{+ \nu} (0, y^-, \mbox{\boldmath $0$}) \: F_\nu^+  
(0, 0, \mbox{\boldmath    $0$}) | P \rangle \;,
\ee        
where $F^{\mu \nu}$ is the gluon field strength tensor and where the summation over         
the colour label has been suppressed.  Due to Bose symmetry we have        
\be        
\label{eq:a7}        
H_g (x) \; = \; H_g (- x) \;.        
\ee

\section{Off-diagonal distributions}   
  
The distributions $H_q $ introduced in (\ref{eq:a2}) may be generalized to allow for  
matrix elements which are off-diagonal in the four momentum of the proton  
\cite{JI}--\cite{JI2}  
\be  
\label{eq:a8}  
H_q (x, \xi, t) \; = \; \frac{1}{2} \: \int \: \frac{dy^-}{2 \pi} \: e^ 
{- i x {\bar P}^+ y^-} \:    
\langle P^\prime | \bar{\psi}_q (0, y^{-}/2, \mbox{\boldmath $0$}) \:          
\textstyle{\frac{1}{2}} \gamma^+ \psi_q (0, -y^{-}/2, \mbox{\boldmath $0$}) | P \rangle  \;,  
\ee    
where we consider only the distributions which conserve  
the proton helicity and which    
describe unpolarized quarks.  Since $\Delta \equiv P - P^\prime \neq 0$ the    
distribution $H_q (x, \xi, t)$ now contains two extra scalar variables,  
in addition to the    
Bjorken $x$ variable.  The variable $t$ is the usual $t$-channel  
invariant, $t = \Delta^2$, and the variable $\xi$ is defined by   
\be   
\label{eq:a9}   
\xi \bar{P}^+ \; = \; \frac{1}{2} \Delta^{+} \;,   
\ee   
where $\bar{P} = \frac{1}{2} (P + P^\prime)$.  This choice of              
variables\footnote{Note that Ji defines $\Delta = P^\prime - P$.} is due to Ji    
\cite{JI}--\cite{JI2}  
and enables symmetry to be imposed between the incoming and outgoing    
proton.  That is Ji uses the symmetric combination $\bar{P}$ of their momenta     
as the defining direction, and calls the $H_q$ off-forward distributions.  The 
distributions $H_q$ are real, and the symmetric choice of  variables has the 
considerable advantage that, due to time-reversal invariance and hermiticity, the 
distributions are even functions of $\xi$  \cite{JI2} 
\be  
\label{eq:a10}  
H_q (x, \xi, t) \; = \; H_q (x, - \xi, t)\;.  
\ee  
Since we will perform our analysis for fixed $t$, concentrating on the
$x$ and $\xi$ dependence,  
we shall omit the $t$ dependence from now on. 
    
To see the physical content of the off-diagonal distributions $H_q$ we again Fourier     
expand $\psi$ and $\bar{\psi}$ in terms of the quark creation and annihilation     
operators.  Since the distributions are even in $\xi$ we may take $\xi > 0$.  In this way     
we obtain the generalization of eq. (\ref{eq:b2}) \cite{JI2}         
\bea   
\label{eq:a11}   
H_q (x, \xi) & = & \frac{1}{2 \bar{P}^+} \int  \frac{d^2 k_T}{2 \sqrt{| 
x^2 - \xi^2    
|} (2 \pi)^3}  \\    
& & \nonumber \\    
& & \sum_\lambda \left [ \langle P^\prime | b_\lambda^\dagger \, ((x - \xi)    
\bar{P}^+, \mbox{\boldmath $k$}_T - \mbox{\boldmath $\Delta$}_T) \: b_\lambda    
\, ((x + \xi) \bar{P}^+, \mbox{\boldmath $k$}_T) | P \rangle \:  
\theta (x \ge \xi) \right .    
\nonumber \\   
& & \nonumber \\  
& & + \; \langle P^\prime | d_\lambda \, ((-x + \xi)  
\bar{P}^+, -\mbox{\boldmath   $k$}_T + \mbox{\boldmath $\Delta$}_T)  
\: b_{- \lambda} \, ((x + \xi) \bar{P}^+,    
\mbox{\boldmath $k$}_T) | P \rangle \: \theta (-\xi < x < \xi) \nonumber  
\\   
& & \nonumber \\  
& & - \; \left . \langle P^\prime | d_\lambda^\dagger \, ((-x - \xi) \bar 
{P}^+,    
\mbox{\boldmath $k$}_T - \mbox{\boldmath $\Delta$}_T) \: d_\lambda \,  
((-x + \xi)    
\bar{P}^+, \mbox{\boldmath $k$}_T) | P \rangle \: \theta (x \le -\xi)  
\right ].    
\nonumber   
\eea   
Fig.~2 gives a pictorial description of the content of (\ref{eq:a11}).  Diagrams (a) and   
(c), which arise from the $b^\dagger b$ and $d^\dagger d$ terms in $\bar{\psi} \psi$,   
generalize Figs.~1(a) and (b) respectively.  For example the first diagram corresponds   
to the emission of a quark of momentum $k$ from the proton followed by its 
absorption with  momentum $k - \Delta$.  Thus for $x > \xi$ and $x < - \xi$  
the off-diagonal   
distribution $H_q$ generalizes the familiar quark and antiquark distributions and will   
evolve according to modified DGLAP equations.  Diagram (b), corresponding to the    
middle region, $- \xi < x < \xi$, does not have a counterpart in Fig.~1.   
This diagram, which   arises from the $db$ term in $\bar{\psi} \psi$,    
corresponds to the emission of a quark-antiquark pair.  In this             
region $H_q$ is a generalization of the proton form factor and will evolve according    
to modified ERBL equations \cite{ERBL}.  Thus in this domain $H_q$ may be    
regarded as a generalization of the  
probability distribution amplitude which occurs in hard exclusive   processes.  
  
Just as for the diagonal case, we introduce valence and singlet quark distributions 
analogous to (\ref{eq:a4})  
\be  
\label{eq:a12}   
H_q^V (x, \xi)  \equiv  H_q (x, \xi) \: + \: H_q (-x, \xi) \; = \; H_ 
q^V (-x, \xi)\;, 
\ee
\be 
H^S (x, \xi)  \equiv  \sum_q \left [H_q (x, \xi) \: - \: H_q (-x, \xi)  
\right ] \; = \; - H^S (-x, \xi)\;.  
\ee  
Thus in addition to the symmetry under $\xi \rightarrow - \xi$, the  
distributions have 
symmetry or antisymmetry under $x \rightarrow -x$.  Also, in analogy to  
(\ref{eq:a7}), the off-diagonal gluon distribution satisfies  
\be  
\label{eq:b12}  
H_g (x, \xi) \; = \; H_g (-x, \xi)\;.  
\ee  
The distributions (\ref{eq:a12})-(\ref{eq:b12}) are identical to those introduced by Ji  
\cite{JI3,JI2}\footnote{Note that  in    
going from Ref.~\cite{JI3} to Ref.~\cite{JI2}  Ji has redefined  
$\xi/2$ by $\xi$.}   except that   
\be   
\label{eq:a13}   
H_g (x, \xi) \; = \; x H_g^{\rm Ji} (x, \xi).   
\ee  
On account of the extra factor $x$, the  gluon distribution (\ref{eq:a13}) is not required 
to be zero at $x=0$, unlike the situation for $H_g^{\rm Ji}$ 
(see also \cite{RAD2} for a relevant discussion).


\section{Off-diagonal distributions on the interval $[0,1]$} 
 
So far we have considered the off-diagonal distributions $H_q (x, \xi)$,  
introduced by    
Ji \cite{JI,JI3},  and  defined on the interval $-1 < x < 1$.   
As noted above $P +P^\prime$ is    
taken as the defining direction, so that symmetry is imposed between the incoming    
$(P)$ and outgoing $(P^\prime)$ proton momenta.  This variable $\xi$ was defined in  
(\ref{eq:a9}) by   
\be   
\label{eq:a15}   
\Delta \; \equiv \; (P - P^\prime) \; = \; \xi (P + P^\prime)\;, 
\ee   
where for simplicity we have omitted the light-cone plus superscript  
(see (\ref{eq:a9})).        
 
To make direct contact with conventional partons we may introduce alternative   
off-diagonal distributions $\hat{\cal{F}}_q (X, \zeta)$ defined on the interval $0 \le   
X \le 1$ such that the initial parton carries a positive fraction $X$ of the proton's         
longitudinal momentum.  That is we take $P$ as the defining direction.  Thus the         
counterpart to (\ref{eq:a15}) is  
\be  
\label{eq:a16}  
\Delta \; = \; \zeta P  
\ee  
with $0 \le \zeta \le 1$.  This is exactly analogous to the approach  
introduced by Radyushkin \cite{RAD,RAD2} in the construction of the non-forward 
distributions ${\cal{F}}_{\zeta}(X)$.  However our construction of  
the distributions $\hat{\cal{F}}_q (X, \zeta)$ presented below 
is different to that of \cite{RAD,RAD2}.  From (\ref{eq:a15}) and  
(\ref{eq:a16}) it follows that  
\be  
\label{eq:a17}  
\xi \; = \; \frac{\zeta}{2 - \zeta}\;.  
\ee  
\subsection{The relation between the distributions $H$ and  $\hat{\cal{F}}$}

In this subsection we first define the off-diagonal distributions  
$\hat{\cal{F}}_q (X, \zeta)$ with $X$ in the interval [0,1]  
starting from Ji's distributions 
$H_q (x, \xi)$  with $x$ in the range [-1,1].   
Then we explore the symmetry relations satisfied  
by the $\hat{\cal{F}}_q (X, \zeta)$. 
  
If we compare the momentum fraction carried by the emitted parton in Fig.~3  
with   
those in Figs.~2(a) and 2(c), then we see that two different  
transformations are         
relevant in reducing the interval $-1 \le x \le 1$ covered by  
$H_q (x,\xi)$ to the         
interval $0 \le X \le 1$ covered by $\hat{\cal{F}}_q (X, \zeta)$.  
 First, from  Fig.~2(a), we have the transformation  
\be  
\label{eq:a18}  
X_1 \; = \; \frac{x_1 + \xi}{1 + \xi}\;,  
\ee  
which takes the interval $x_1\in [-\xi,1]$ into $X_1\in [0,1]$. Simultaneously 
$\xi$ is transformed into $\zeta$.  
Secondly, from  Fig.~2(c), we have the transformation  
\be  
\label{eq:a19}  
X_2 \; = \; \frac{\xi - x_2}{1 + \xi}\;,  
\ee  
which takes $x_2\in [-1,\xi]$ into $X_2\in [0,1]$.  
Now, $-\xi$ is transformed into $\zeta$.  
In this way we introduce two   
distinct off-diagonal distributions $\hat{\cal{F}}_q$ and               
\bea  
\label{eq:a}  
\hat{\cal{F}}_q (X_1, \zeta) & = & \frac{1}{1 - \zeta/2} \;   
H_q (x_1, \xi) \nonumber \\  
& & \\  
\hat{\cal{F}}_{\bar{q}} (X_2, \zeta) & = & \frac{-1}{1 -\zeta/2} \; H_q (x_2, 
\xi)\;, \nonumber  
\eea  
where $\xi = \zeta/(2 - \zeta)$ and  the inverse relations
\be  
\label{eq:a21}  
x_1 \; = \; \frac{X_1 - \zeta/2}{1 - \zeta/2}\;, \quad\quad\quad x_2 \; = \;  
\frac{\zeta/2 - X_2}{1 -  \zeta/2}  
\ee  
follow from (\ref{eq:a17}--\ref{eq:a19}).  We stress that as $X_{1,2}$ cover the   
range $[0,1]$, the corresponding $x_1$ and $x_2$ cover respectively the ranges  
$[-\xi,  1]$ and $[-1, \xi]$, as shown schematically in Fig.~4.   
The factors  
$\pm (1 -\zeta/2)^{-1}$ in (\ref{eq:a}) arise from the translation of the measure   
$dx$ to $dX$.  
  
In the limit that $\zeta$ (and $\xi$) $\rightarrow 0$ we have from  
(\ref{eq:a3})  
\bea  
\label{eq:a22}  
\hat{\cal{F}}_q (X, 0) & = & ~~~~H_q (X, 0) \; = \; q (X) \nonumber \\  
& & \\  
\hat{\cal{F}}_{\bar{q}} (X, 0) & = & -H_q (-X, 0) \; = \;  
\bar{q}(X)\;,   
\nonumber  
\eea  
which is an additional motivation for using the quark and antiquark subscripts to  
differentiate between the two  functions $\hat{\cal{F}}_q$ and   
$\hat{\cal{F}}_{\bar{q}}$.  
  
Finally, due to the symmetry relation (\ref{eq:b12}), the gluon distribution may be         
defined in the range $0 \le X \le 1$ by either of the transformations (\ref{eq:a21}).          
That is we have        
\be  
\label{eq:a23}  
\hat{\cal{F}}_g (X, \zeta) \; = \; \frac{1}{1 - \zeta/2} \: H_g   
\left ( \frac{X - \zeta/2}{1 - \zeta/2}, \xi   
\right ) \; = \; \frac{1}{1 - \zeta/2} \:  
H_g \left (   
\frac{\zeta/2 - X}{1 - \zeta/2}, \xi \right ).  
\ee

\subsection{Symmetry relations}

From Fig.~4 we see that in the DGLAP-type regions ($x > \xi$ or $x < - \xi$) $H_q$   
is transformed respectively into {\it independent} functions $\hat{\cal{F}}_q (X)$         
and $\hat{\cal{F}}_{\bar{q}} (X)$ with $X > \zeta$.  On the other hand in the         
ERBL-type region ($- \xi < x < \xi$) the distribution $H_q$ generates functions         
$\hat{\cal{F}}_q (X)$ and $\hat{\cal{F}}_{\bar{q}} (X)$ 
with $X < \zeta$ which are no longer independent.  Indeed for $X < \zeta$ we have                
\bea  
\label{eq:a24}  
\hat{\cal{F}}_q (\zeta - X) & = & \frac{1}{1 -\zeta/2} \: H_q   
\left ( \frac{\zeta - X - \zeta/2}{1 - \zeta/2}   
\right ) \; = \; \frac{1}{1 - \zeta/2} \;  
H_q \left (  \frac{\zeta/2 - X} 
{1 - \zeta/2} \right )  
\nonumber \\ 
\nonumber \\  
&=&    
\; - \hat{\cal{F}}_{\bar{q}} (X)\;,  
\eea          
where for simplicity we do not indicate the  additional explicit 
$\zeta$ or $\xi$ dependence of the  distributions.  

Eq.~(\ref{eq:a24}) is the basic symmetry 
relation for the off-diagonal quark distributions  
which indicates that in the ERBL-like region the
quark and antiquark distributions are not independent,
unlike the case in the DGLAP-like
region. The  physical reason for this can easily be  
understood by looking at Fig.~2b. In the ERBL-like  
region we can define the off-diagonal 
distributions with respect to the first emitted parton being 
either the quark with momentum 
$x+\xi$ or the antiquark with momentum $\xi-x$. The latter possibility corresponds 
to the exchange of the annihilation operators in eq. (\ref{eq:a11}), 
which is the origin of  the $-$ sign in relation (\ref{eq:a24}). 
         
We may form the non-singlet and singlet combinations of the quark and antiquark          
distributions.  From (\ref{eq:a12}) we have          
\bea         
\label{eq:a25}         
\hat{\cal{F}}_q^V (X) & = & \frac{1}{1 - \zeta/2} \; H_q^V          
\left (\frac{X - \zeta/2}{1 - \zeta/2} \right )          
\; = \; \hat{\cal{F}}_q (X) - \hat{\cal{F}}_{\bar{q}} (X)\;,  
\nonumber \\         
& & \\         
\hat{\cal{F}}^S (X) & = & \frac{1}{1 - \zeta/2} 
 \; H^S \left (          
\frac{X - \zeta/2}{1 - \zeta/2} \right )  
\; = \; \sum_q \left [ \hat{\cal{F}}_q (X) + \hat{\cal{F}}_{\bar{q}} (X)  
\right ]\;,          
\nonumber         
\eea         
which in the region $X < \zeta$ satisfy symmetry relations resulting from  
(\ref{eq:a24})        
\bea         
\label{eq:a26}         
\hat{\cal{F}}_q^V (\zeta - X) & = & \hat{\cal{F}}_q^V (X)\;,  
\nonumber \\         
& & \\         
\hat{\cal{F}}^S (\zeta - X) & = & - \hat{\cal{F}}^S (X)\;.  
\nonumber  
\eea         
It is straightforward to show for $X < \zeta$ that the gluon distribution 
(\ref{eq:a23}) satisfies a similar relation        
\be         
\label{eq:a27}         
\hat{\cal{F}}_g (\zeta - X) \; = \; \hat{\cal{F}}_g (X)\;.              
\ee                  
These properties are well-illustrated by Fig.~5.  The upper plot shows an example of 
the            
off-diagonal distribution $H_q (x, \xi)$ with $x\in [-1,1]$, for $\xi = 0.5$.   
The middle plot shows the transformation of this distribution into the two          
functions $\hat{\cal{F}}_q (X, \zeta)$ and $\hat{\cal{F}}_{\bar{q}} (X, \zeta)$ of          
(\ref{eq:a}) with $X\in [0,1]$.   
Their behaviour shows that the symmetry relation (\ref{eq:a24}) is          
clearly satisfied in the region $0 \le X \le \zeta$.  Finally, the lower plot shows the          
behaviour of the non-singlet $\hat{\cal{F}}_q^V$ and  
the singlet  $\hat{\cal{F}}^S$ combinations.   
The symmetry of $\hat{\cal{F}}_q^V$  and          
antisymmetry of $\hat{\cal{F}}^S$, about the point  
$X = \zeta/2$, are   clearly evident in the region $0 \le X \le \zeta$.

\section{Evolution equations}        
 
Just as we constructed $\hat{\cal{F}}$ directly 
from the  off-forward distributions $H$ of Ji, so       
we start with the evolution equations \cite{JI3} for $H(x, \xi)$  
with $x\in [-1,1]$  
and use transformations (\ref{eq:a}) and (\ref{eq:a23}) to rewrite them in        
terms of the distributions $\hat{\cal{F}}(X, \zeta)$ with $X\in [0,1]$. 
   
In  the DGLAP-like region $X > \zeta$ (which corresponds to $x > \xi$ or $x < -\xi$) 
the equations that we obtain for $\hat{\cal{F}}$        
are equivalent to those given for the non-forward distributions of Radyushkin 
\cite{RAD2,GKM}.  
Their full form can  be found in the Appendix. 
Moreover in the limit $\zeta \rightarrow 0$  
they reduce to the familiar DGLAP evolution equations.  
            
However in the ERBL-like region $X < \zeta$ (corresponding to $-\xi < x  < \xi$) 
the equations obtained for        
$\hat{\cal{F}}$ are different to those given in \cite{RAD2,GKM} for the non-
forward 
distributions. They have the following forms       
\bea        
\label{eq:a28}        
\mu \frac{\partial}{\partial \mu} \: \hat{\cal{F}}_q^V (X, \zeta) & = &  
P_{QQ}\otimes \hat{\cal{F}}_q^V   
\nonumber \\ 
\nonumber \\  
&+& \frac{\alpha_S C_F}{\pi}  
\int_\zeta^1        \frac{dZ}{Z} \! \left [ \frac{Z}{X - \zeta + Z} -  
\frac{X}{\zeta} \right ] \:         
\hat{\cal{F}}_q^V (Z, \zeta) \\        
& & \nonumber \\        
& & \nonumber \\  
\label{eq:a29}\nonumber      
\mu \frac{\partial}{\partial \mu} \hat{\cal{F}}^S (X, \zeta) & = &  
P_{QQ} \otimes \hat{\cal{F}}^S \: + \: P_{QG} \otimes \hat{\cal{F}}_g  
\nonumber \\  
& -&\frac{\alpha_SC_F}{\pi} \int_\zeta^1 \frac{dZ}{Z}  
\! \left [ \frac{Z}{X - \zeta + Z} - \frac{X}{\zeta} \right ] \:  
\hat{\cal{F}}^S (Z, \zeta)  
\\  \nonumber     
& & \\  \nonumber      
& + &\: \frac{\alpha_S N_f}{\pi} \int_\zeta^1 \frac{dZ}{Z} \;  
\frac{(1 - \zeta/2)(\zeta - X)}{\zeta^2} \; \left [  
\frac{4 X}{\zeta} +         
\frac{2X - \zeta}{Z} \right ] \:  
\hat{\cal{F}}_g (Z, \zeta)  \nonumber \\        
& & \nonumber \\        
& & \nonumber \\       
\label{eq:a30}      
\mu \frac{\partial}{\partial \mu} \hat{\cal{F}}_g (X, \zeta) & = &  
P_{GQ} \otimes \hat{\cal{F}}^S \:+ \: P_{GG} \otimes \hat{\cal{F}}_g  
\nonumber \\        
\nonumber \\    
& - &\frac{\alpha_S C_F}{\pi} \int_\zeta^1 \frac{dZ}{Z}  
\frac{(\zeta - X)^2}{\zeta (1 -\zeta/2)}  
\: \hat{\cal{F}}^S (Z, \zeta) \\  \nonumber       
& & \\ \nonumber        
& +& \: \frac{\alpha_S N_c}{\pi} \int_\zeta^1 \frac{dZ}{Z}  
\frac{(\zeta - X)^2}{Z} \! \left [ \frac{1}{X - \zeta + Z} + \frac{2Z}{\zeta^2}  
\left (1 + \frac{2X}{\zeta} +         
\frac{X}{Z} \right ) \right ] \: \hat{\cal{F}}_g (Z, \zeta)\;,        
\eea       
where the scale $\mu$ is implicit in the distributions $\hat{\cal{F}}$. 
The full  forms of the equations are given in the Appendix.  Here it is sufficient  
to note that the  convolutions shown symbolically as $P \otimes \hat{\cal{F}}$  
are identical to those given in \cite{RAD2,GKM}.  
However the new evolution equations contain several additional terms, each        
being a convolution integral over the range $[\zeta, 1]$.  These extra terms are        
essential to preserve the symmetry properties (\ref{eq:a26}) and (\ref{eq:a27})  
of $\hat{\cal{F}}$ during the evolution. We note that in the limit  
$\zeta \rightarrow 1$ 
the additional terms are to 
equal zero and that (\ref{eq:a28}) - (\ref{eq:a30})  reduce to 
the ERBL evolution equations \cite{ERBL} for the distribution amplitudes.

\subsection{Numerical results of the evolution} 
 
To illustrate how the off-diagonal distributions $\hat{\cal{F}}$ evolve  
with increasing 
renormalization scale $\mu$ 
we constructed a computer programme based on the equations given  
in the Appendix. 
For the initial input  
at the starting scale $\mu = 1$~GeV we adopt the following strategy.  
 We start with given input forms for the  
off-forward distributions $H_{q,g} (x, \xi)$, which are even in $\xi$. 
An  example for the quark distribution is shown in Fig.~5(a).  
Then using prescriptions (\ref{eq:a}) and (\ref{eq:a23}) we  
transform $H_{q,g} (x, \xi)$ into the distributions  
$\hat{\cal{F}}_{q, \bar{q},g}(X, \zeta)$  
which satisfy the symmetry relations (\ref{eq:a26}) and (\ref{eq:a27}).   
The initial distribution $H_{q} (x, \xi)$ shown in Fig.~5(a) is only  meant to illustrate 
the general features of the adopted strategy. The detailed properties of more realistic 
initial distributions will be discussed in a separate paper. 
 
The results that are obtained by evolving  
$\hat{\cal{F}}^V$, $\hat{\cal{F}}^S$ and $\hat{\cal{F}}_g$  
to higher scales are  
shown in the three plots of Fig.~6.  In each plot the dashed curve  
is the input at $\mu = 1$~GeV, while the dot-dashed curve shows  
the effect of evolution up to $\mu = 10$~GeV.   
It is evident that evolution does indeed preserve the symmetry  
properties in the ERBL-like region, $X < \zeta$.   
 
The continuous curves in Fig.~6 are the results  
of evolving all the way up to $\mu~\rightarrow~\infty$.   
These asymptotic forms are identical 
with the analytic asymptotic solutions \cite{RAD2,RAD3} of the evolution equations  
for the distributions $\ftil$ given in the Appendix 
\bea    
\label{eq:a33}    
\hat{\cal{F}}_q^V (X, \zeta) & \sim & \frac{X}{\zeta}  
\left (1 - \frac{X}{\zeta} \right ) \nonumber \\    
& & \nonumber \\    
\hat{\cal{F}}^S (X, \zeta) & \sim & \frac{X}{\zeta}  
\left (1 - \frac{X}{\zeta} \right )     
\left ( \frac{2X}{\zeta} - 1 \right ) \\    
& & \nonumber \\    
\hat{\cal{F}}_g (X, \zeta) & \sim & \left ( \frac{X}{\zeta} \right )^2  
\left (1 -\frac{X}{\zeta} \right )^2. \nonumber    
\eea     
A remarkable property \cite{RAD2} is evident from Fig.~6.  We see that the  
distributions are swept from the DGLAP-like to the ERBL-like region as $\mu$  
increases.  Indeed the asymptotic forms show that  
they are finally entirely contained in the ERBL-like region with  
$X < \zeta$.

\section{Relation to the non-forward distributions} 
 
The off-diagonal distributions $\hat{\cal{F}} (X, \zeta)$,   
constructed in the previous section, 
are equivalent to the 
off-forward distributions $H(x,\xi)$ defined by Ji. They are also closely 
related to, but not the same as, the non-forward distributions 
${\cal{F}}_\zeta (X)$ introduced by Radyushkin\footnote{We thank  
A.V.~Radyushkin for helpful comments on the subject of this section.}.
The difference between  them occurs in the
ERBL-like region $(X<\zeta)$.

The non-forward distributions ${\cal{F}}^{q,\qbar}_\zeta (X)$ are related to  
the off-forward distributions $H_q(x,\xi)$ in the following way 
(see  Section IX of Ref.~\cite{RAD2} for a detailed discussion)
\bea 
\label{eq:61} 
 (1+\xi)~H_q(x,\xi)\;=\; 
\left\{  
\begin{array}{ll} 
 {\cal{F}}^{q}_\zeta(X)                                & \mbox{if  $x>\xi$}  \\ 
 \\ 
 {\cal{F}}^{q}_\zeta(X)-{\cal{F}}^{\qbar}_\zeta(\zeta-X)  & \mbox{if  $-
\xi<x<\xi$}  \\ 
 \\ 
 -{\cal{F}}^{\qbar}_\zeta(\zeta-X)      & \mbox{if  $x<-\xi$}\;,  
\end{array} 
\right. 
\eea 
where $X=(x+\xi)/(1+\xi)$ and $\zeta=2\xi/(1+\xi)$. Notice that while in 
the DGLAP-like regions 
($x>\xi$ or $x<-\xi$) there is a one-to-one correspondence between the two 
distributions, 
in the ERBL-like region ($-\xi<x<\xi$) Ji's distribution $H_q$ only determines a 
specific 
combination of Radyushkin's distributions ${\cal{F}}^{q,\qbar}$. This is in contrast 
to the
distributions defined by eqs.~(\ref{eq:a}) which are in  one-to-one correspondence 
with $H_q$. 
 
Comparing eqs.~(\ref{eq:61}) with eqs.~(\ref{eq:a}) we see that the off-diagonal 
distributions 
$\ftil$ are identical to the non-forward distributions ${\cal{F}}$ in the DGLAP-like 
region 
$(X>\zeta)$. However in the ERBL-like region there are different.   
To be precise, for $X < \zeta$, we have 
\bea 
\label{eq:b27} 
\hat{\cal{F}}_q (X, \zeta) & = & {\cal{F}}_\zeta^q (X) \; - \;  
{\cal{F}}_\zeta^{\bar{q}} (\zeta - X) \nonumber \\ 
\\ 
\hat{\cal{F}}_{\bar{q}} (X, \zeta) & = & {\cal{F}}_\zeta^{\bar{q}} (X) \; - \;  
{\cal{F}}_\zeta^q (\zeta - X)\;. \nonumber 
\eea 
The main difference between the distributions $\hat{\cal{F}} (X,\zeta)$ 
and the non-forward distributions ${\cal{F}}_\zeta (X)$ is that the latter 
do not obey the symmetry properties (\ref{eq:a24}) and (\ref{eq:a26})-(\ref{eq:a27}). 
These properties are essential for our distributions. 
They result from the construction which ensures the equivalence of the distributions $\ftil$
to Ji's distributions $H$. The physical reason for the symmetries was discussed
in Section~4.2. 
An important consequence of the symmetry relations is that in the ERBL-like region
the quark and antiquark off-diagonal distributions are not independent, 
see relation (\ref{eq:a24}). 

This should be contrasted to the case of the
non-forward distributions  of Radyushkin. They are are obtained through the integration
of ``double distributions'' $F$ which are universal $\zeta$-independent functions. 
The  double distributions are separated into two 
independent components (which are denoted by $F_q$ and $F_{\bar q}$)
according to the sign of $x$ in the exponential.  As a result the corresponding
non-forward distributions ${\cal{F}}^{q}$ and ${\cal{F}}^{\bar q}$
are also independent in the ERBL-like region, 
see \cite{RAD3} for more details. 
Thus there are twice as many quark ``degrees of freedom'' in the ERBL-like region 
as in our case.

A similar comparison can be done for the non-singlet, singlet and gluon distributions. 
As a result we find the following relations for $X<\zeta$ 
\bea 
\label{eq:b28} 
\hat{\cal{F}}^V (X, \zeta) & = & {\cal{F}}_\zeta^{V}(X) + 
{\cal{F}}_\zeta^{V}(\zeta-X) 
\nonumber \\ 
\nonumber \\  
\hat{\cal{F}}^S (X, \zeta) & = & {\cal{F}}_\zeta^{S}(X) - 
{\cal{F}}_\zeta^{S}(\zeta-X) 
\\ 
\nonumber \\ 
\hat{\cal{F}}^g (X, \zeta) & = & {\cal{F}}_\zeta^{g}(X) + 
{\cal{F}}_\zeta^{g}(\zeta-X)\;. 
\nonumber 
\eea 
Thus we see that the distributions $\hat{\cal{F}}$ 
are equal to symmetric or antisymmetric combinations 
of the corresponding non-forward distributions ${\cal{F}}$ 
in the ERBL-like region.  These combinations of the non-forward distributions
were used in \cite{RAD2} in the description of the ERBL-like region.
However the further analysis in Ref.~\cite{RAD2} was done in terms of the 
unsymmetrized  non-forward distributions ${\cal{F}}_{\zeta}^{(V,S,g)}(X)$.

\subsection{Comparison of the two sets of evolution equations} 
     
The evolution equations for the non-forward distributions  
${\cal{F}}^{(V,S,g)}$ of Refs.~\cite{RAD2,GKM}  
do not obey the symmetry properties (\ref{eq:a26}) and (\ref{eq:a27}) in the ERBL-
like region. 
One may try, however, to write down the evolution equations for the combinations  
on the right hand side of eqs.~(\ref{eq:b28}), starting  from the equations given in  
\cite{RAD2,GKM} for the full non-forward distributions   
\bea 
\label{eq:b30} 
{\cal{F}}_\zeta (X) & \equiv & {\cal{F}}_\zeta^{({\rm sym})} (X) \: + \:  
{\cal{F}}_\zeta^{({\rm asym})} (X) \nonumber \\ 
& & \\ 
& = &  
\textstyle{\frac{1}{2}} \left [{\cal{F}}_\zeta (X) +  
{\cal{F}}_\zeta (\zeta - X)  
\right ] \: + \: \textstyle{\frac{1}{2}} \left [ {\cal{F}}_\zeta (X) -  
{\cal{F}}_\zeta (\zeta - X) \right ] \nonumber 
\eea 
Not surprisingly these ``symmetrized'' evolution equations are almost identical to  
the evolution equations for the distributions (\ref{eq:a28})--(\ref{eq:a30}). 
The integrals over $[\zeta,1]$, indicated explicitly in (\ref{eq:a28})--(\ref{eq:a30}),  
appear in the ``symmetrized'' equations
as a result of the symmetrization procedure. The only difference 
appears in the ``symmetrized'' gluon equation which additionally contains a term 
proportional to the integral over the full non-forward singlet distribution 
\be      
\label{eq:a32}      
\int_0^1  dZ \; {\cal{F}}_{\zeta}^S (Z) \; = \; \int_0^\zeta dZ \; 
{\cal{F}}_\zeta^{S ({\rm sym})} (Z) 
\: + \: \int_\zeta^1 dZ\; {\cal{F}}_\zeta^S (Z)   \;.      
\ee      
Since the ``symmetrized'' gluon equation should contain only the asymmetric singlet  
combination ${\cal{F}}_\zeta^{S ({\rm asym})}$, the above term mixes the 
symmetric and antisymmetric components of the singlet distribution, and thus violates 
the symmetry properties (\ref{eq:a26}) and (\ref{eq:a27}) for the singlet and gluon 
distributions. The only case when 
it does not happen is if the integral ({\ref{eq:a32}) is equal to zero due to initial 
conditions. 
The value of this integral is conserved by the evolution equations 
\cite{RAD2,GKM} for the full non-forward 
distributions\footnote{We were informed by A.V. Radyushkin that
the above mentioned problem with the integral (\ref{eq:a32}) 
can be solved if one uses the kernel $P_{GQ}$ in the evolution equations
of Ref.~\cite{RAD2} in the form originally obtained by Chase \cite{CH}.
The method used in Ref.~\cite{RAD2} cannot unambiguously fix this kernel.}. 
Only in this case may  
the non-forward distributions of Radyushkin be equivalent to the  
off-diagonal distributions of Ji. This can be done by taking into account only 
one of the two parts in the decomposition (\ref{eq:b30}) --- symmetric for the non-
singlet 
and gluon, and antisymmetric for the singlet, distributions.

\subsection{The singularity structure of the basic amplitude} 
  
For the purpose of illustration we may consider the classic process of  
deeply virtual  Compton scattering.   
The invariant amplitude for the process has the generic form  \cite{JI3}  
\be  
\label{eq:a34}  
T \; \sim \; \int_{-1}^1 dx \; \left [ \frac{1}{x - \xi  + i \varepsilon} 
 \: + \: \frac{1}{x +   
\xi - i \varepsilon} \right ] \; H_q (x, \xi).  
\ee  
If the amplitude is translated into a form involving the distributions     
$\hat{\cal{F}}_q (X, \zeta)$ defined on the interval $0 \le X \le 1$,  
then  (\ref{eq:a34}) becomes  
\be  
\label{eq:a35}  
T \; \sim \;  
\int_0^1 dX \;   
\frac{\hat{\cal{F}}_q (X, \zeta) + \hat{\cal{F}}_{\bar{q}} (X, \zeta)} 
{X - \zeta + i \varepsilon}  
\: + \: \int_\zeta^1 \frac{dX}{X} \; \left [\hat{\cal{F}}_q (X, \zeta) +   
\hat{\cal{F}}_{\bar{q}} (X, \zeta) \right ]\;.   
\ee 
We see that (\ref{eq:a35}) contains only one singularity at $X = \zeta$, which results  
from the quark propagator, and is regularized by the $+ i \varepsilon$ prescription
and assuming that  $\hat{\cal{F}}_{q,\bar q} (X, \zeta)$ are continuous at $X=\zeta$. 
Note that there is no singularity at $X = 0$ since the second integral is bounded by  
$\zeta > 0$ from below.   
 
This is in contrast to the amplitude derived using the non-forward distributions 
${\cal{F}}_\zeta^q (X)$ 
\cite{RAD,RAD2}. Then $T$ contains a second singularity at $X = 0$, since in this 
case   
the second integral in (\ref{eq:a35}) goes from 0 to 1. The result can be  
derived by substituting relations (\ref{eq:b27}) into (\ref{eq:a35}). 
This additional (end-point) singularity is removed  
by assuming that the non-forward distributions 
${\cal{F}}_\zeta^{q,\bar q}(X)$ vanish as $X \rightarrow 0$. Looking at
eq.~(\ref{eq:61}) we see that this assumption is equivalent to 
the continuity of $H(x,\xi)$ at $x=\pm \xi$ 
(or $\ftil(X,\zeta)$ at $X=\zeta$)\footnote{We thank A.V. Radyushkin for this remark}.
Such  assumption 
is not required for our off-diagonal distributions $\ftil(X,\zeta)$, see eq.~(\ref{eq:a35}).   
Indeed, if present, it would clearly  
violate their continuity at $X = \zeta$,  
or their symmetry about $X=\zeta/2$,  
see Fig.~5.

\section{ Conclusions}

In this paper we have transformed the off-forward parton distributions  
$H (x, \xi)$ defined by Ji, in which the defining direction is the average between 
the incoming and outgoing proton momenta and $x\in [-1,1]$,   
into off-diagonal distributions $\hat{\cal{F}} (X,\zeta)$,  
in which the defining direction is the    
incoming proton momentum  and $X\in [0,1]$.     
These off-diagonal distributions $\hat{\cal{F}} (X,\zeta)$ therefore have a close  
identification with conventional (diagonal) distributions.   
Moreover, by construction, they are fully 
equivalent to the off-forward distributions of Ji. 
 
In the ERBL-like domain $(X<\zeta)$ they satisfy the symmetry relations   
\be   
\label{eq:a36}   
\hat{\cal{F}} (\zeta - X, \zeta) \; = \; \pm \hat{\cal{F}} (X, \zeta)   
\ee   
where the $+$ sign applies to the gluon and quark non-singlet distributions, and the    
$-$ sign applies to the quark singlet.  We presented the evolution equations satisfied  
by the $\hat{\cal{F}} (X, \zeta)$ and gave numerical results (Figs.~5 and 6) to  
illustrate the properties of the distributions.  We found that asymptotically $(\mu  
\rightarrow \infty)$ the distributions evolve to the known analytic asymptotic forms.   
Indeed as $\mu$ increases the distributions are swept from the DGLAP-like domain to 
lie  
entirely within the ERBL-like region, as illustrated by the example shown in Fig.~6.   
The symmetry relations (\ref{eq:a36}) are preserved at each stage of the evolution.   
   
The  distributions $\hat{\cal{F}}(X,\zeta)$ are analogous to, but not 
the same as, the non-forward distributions ${\cal{F}}_\zeta (X)$ introduced   
by Radyushkin \cite{RAD}. The difference lies in the ERBL-like region, 
since the non-forward distributions do not obey the symmetry  
relations (\ref{eq:a36}). As a result the non-forward distributions ${\cal{F}}_\zeta 
(X)$ 
are not in general equivalent to the off-forward distributions $H(x,\xi)$ of 
Ji. We stressed that this happens only   in the ERBL-like region.
We discussed conditions under which ${\cal{F}}_\zeta (X)$ would become equivalent
to $H(x,\xi)$ (and $\hat{\cal{F}}(X,\zeta)$).  
We also commented on   the singularity at $X=0$ 
of the basic DVCS amplitude at tree level when written in terms of 
${\cal{F}}_\zeta (X)$, which requires  
${\cal{F}}_\zeta (X)$ to vanish as $X \rightarrow 0$.  
The distributions $\hat{\cal{F}}(X,\zeta)$, which we defined,   
have the advantage that they do not lead to such a singularity. 
   
\bigskip 
 
\noindent {\large \bf Acknowledgements}   
   
We thank Xiangdong Ji and Anatoly Radyushkin for discussions.  K.G.B. thanks the    
Royal Society/NATO and the UK Particle Physics and Astronomy Research Council    
for financial support.  This research has also been supported in part by the Polish State    
Committee for Scientific Research grant No.~2~P03B~089~13 and by the EU Fourth    
Framework Programme \lq Training and Mobility of Researchers' Network, \lq    
Quantum Chromodynamics and the Deep Structure of Elementary Particles', contract    
FMRX-CT98-0194 (DG~12-MIHT).


\newpage 
\section*{Appendix}

Here we present for  reference the full form of the evolution equations
for our  non-singlet $\ftil^V_q(X,\zeta,\mu)$,  
singlet $\ftil^S(X,\zeta,\mu)$ and
gluon $\ftil_g (X,\zeta,\mu)$ distributions 
defined in the range $0\le X\le 1$ by eqs.~(\ref{eq:a25}) 
and (\ref{eq:a23}). The asymmetry parameter $\zeta$ lies in the range
$[0,1]$.

We use the following notation 
$\Xprime \equiv X-\zeta$ and $\Zprime \equiv Z-\zeta$ and suppress 
the renormalization scale $\mu$ among the arguments of our distributions.
In the DGLAP-like region $X>\zeta$
we have the following evolution equations 
\bea
\label{eq:ap1}
\nonumber
\mu\frac{\partial}{\partial \mu} \ftil^V_q(X,\zeta,\mu)
&=&
\frac{\alpha_S}{\pi}  C_F\;
\Biggr\{
\int\limits_X^1 \frac{dZ}{X-Z}\;
\biggl[
\biggl(\frac{X}{Z}+\frac{\Xprime}{\Zprime}\biggr)\;\ftil^V_q(X,\zeta)-
\biggl(1+\frac{X \Xprime}{Z \Zprime}\biggr)\;\ftil^V_q(Z,\zeta)
\biggr]
\nonumber \\
\nonumber \\
\qquad\qquad\quad &+&\;
\ftil^V_q(X,\zeta)\;\biggl[\;\frac{3}{2} + \ln\frac{(1-X)^2}{1-\zeta}\biggr]
\Biggr\}\;,
\nonumber
\eea
\bea
\label{eq:ap2}
\mu\frac{\partial}{\partial \mu}\ftil^S(X,\zeta,\mu)
&=&
\frac{\alpha_S}{\pi}  C_F\;
\Biggr\{
\int\limits_X^1 \frac{dZ}{X-Z}\;
\biggl[
\biggl(\frac{X}{Z}+\frac{\Xprime}{\Zprime}\biggr)\;\ftil^S(X,\zeta)-
\biggl(1+\frac{X \Xprime}{Z \Zprime}\biggr)\;\ftil^S(Z,\zeta)
\biggr]
\nonumber \\
\nonumber \\
\qquad\qquad\quad &+&\;
\ftil^S(X,\zeta)\;\biggl[\;\frac{3}{2} + \ln\frac{(1-X)^2}{1-\zeta}
\biggr]
\Biggr\}
\nonumber \\
\nonumber \\
&+&
\frac{\alpha_S}{\pi} N_f 
\int\limits_X^1 \frac{dZ}{Z \Zprime}\;\biggl(1-\frac{\zeta}{2}\biggr)\;
\biggl[\biggl(1-\frac{X}{Z}\biggr)\biggl(1-\frac{\Xprime}{\Zprime}\biggr)\;
+\frac{X \Xprime}{Z \Zprime}
\;\biggr]\;\ftil_g(Z,\zeta)
\;,
\nonumber 
\eea

\bea
\label{eq:ap3}
\mu\frac{\partial}{\partial \mu}\ftil_g(X,\zeta,\mu)
&=&
\frac{\alpha_S}{\pi} C_F\;
\int\limits_X^1 dZ\;
\biggl[\biggl(1-\frac{X}{Z}\biggr)
\biggl(1-\frac{\Xprime}{\Zprime}\biggr) +1
\biggr]\;\frac{\ftil^S(Z,\zeta)}{1-{\zeta}/{2}}
\nonumber \\
\nonumber \\
&+&
\frac{\alpha_S}{\pi} N_c\;
\Biggr\{
\int\limits_X^1 dZ\; 
\biggl[
\frac{2}{Z} \biggl(1+\frac{X \Xprime}{Z \Zprime}\biggr)
\biggl(1-\frac{\Xprime}{\Zprime}\biggr)\;\ftil_g(Z,\zeta)
\nonumber \\
\nonumber \\ 
&+&
\frac{\bigl[({X}/{Z})+({\Xprime}/{\Zprime})\bigr]\; \ftil_g(X,\zeta)
-\bigl[({X}/{Z})^2+({\Xprime}/{\Zprime})^2\bigr]\; \ftil_g(Z,\zeta)}
{X-Z}
\biggr]
\nonumber \\
\nonumber \\
\qquad\qquad\quad &+&\;
\ftil_g(X,\zeta)\;
\biggl[\frac{11-(2 N_f)/3}{2N_c}+\ln\frac{(1-X)^2}{1-\zeta}
\biggr]
\Biggr\}~,
\eea
where $C_F=4/3$ and $N_c=3$, and $N_f$ is the number of active flavours.
In the limit $\zeta=0$ the above equations become the familiar DGLAP evolution
equations.


The equations in the ERBL-like region $X<\zeta$ 
are more complicated since they
involve integration with different kernels in the intervals
$[0,X]$ and $[X,1]$. We have 

\bea
\label{eq:ap4}
\mu\frac{\partial}{\partial \mu}\ftil^V_q(X,\zeta,\mu)
&=&
\frac{\alpha_S}{\pi}  C_F\;
\Biggr\{
\int\limits_0^X dZ\; \biggl(\frac{\Xprime}{\Zprime} \biggr)
\;\biggl[
\frac{\ftil^V_q(Z,\zeta)}{\zeta}+
\frac{\ftil^V_q(Z,\zeta)-\ftil^S(X,\zeta)}{X-Z}
\biggr]
\nonumber \\
\nonumber \\
\qquad\qquad\quad &+&\;
\int\limits_X^1 dZ\; \biggl( \frac{X}{Z}\biggr)
\biggl[
\frac{\ftil^V_q(Z,\zeta)}{\zeta}+
\frac{\ftil^V_q(Z,\zeta)-\ftil^S(X,\zeta)}{Z-X}
\biggr]
\nonumber \\
\nonumber \\
\qquad\qquad\quad &+&\;
\ftil^V_q(X,\zeta,\mu)\;
\biggl[
\frac{3}{2}+\ln\frac{X(1-X)}{\zeta}
\biggr]
\nonumber \\
\nonumber \\
& &  + \int_\zeta^1 \frac{dZ}{Z} \; \left [\frac{Z}{X - \zeta + Z} - 
\frac{X}{\zeta} \right ] \;        
\tilde{\cal{F}}_q^V (Z,\zeta)
\Biggr\}
\nonumber
\eea
\bea
\label{eq:ap5}
\mu\frac{\partial}{\partial \mu}\ftil^S(X,\zeta,\mu)
&=&
\frac{\alpha_S}{\pi}  C_F\;
\Biggr\{
\int\limits_0^X dZ\; \biggl(\frac{\Xprime}{\Zprime} \biggr)
\;\biggl[
\frac{\ftil^S(Z,\zeta)}{\zeta}+
\frac{\ftil^S(Z,\zeta)-\ftil^S(X,\zeta)}{X-Z}
\biggr]
\nonumber \\
\nonumber \\
\qquad\qquad\quad &+&\;
\int\limits_X^1 dZ\; \biggl( \frac{X}{Z}\biggr)
\biggl[
\frac{\ftil^S(Z,\zeta)}{\zeta}+
\frac{\ftil^S(Z,\zeta)-\ftil^S(X,\zeta)}{Z-X}
\biggr]
\nonumber \\
\nonumber \\
\qquad\qquad\quad &+&\;
\ftil^S(X,\zeta)\;
\biggl[
\frac{3}{2}+\ln\frac{X(1-X)}{\zeta}
\biggr]
\nonumber \\
\nonumber \\
&  & - \int_\zeta^1 \frac{dZ}{Z} \; \left [\frac{Z}{X - \zeta + Z} - 
\frac{X}{\zeta} \right ] \;        
\tilde{\cal{F}}^S (Z,\zeta)
\Biggr\}
\nonumber \\
\nonumber \\
&+& 
\frac{\alpha_S}{\pi} N_f\;
\Biggr\{
\int\limits_0^X \frac{dZ}{\zeta^2}\; \biggl(1-\frac{\zeta}{2}\biggr)\;
\biggl(\frac{\Xprime}{\Zprime} \biggr)
\biggl[
\;4 \frac{X}{\zeta} + \frac{2X-\zeta}{\zeta-Z}
\biggr]\;\ftil_g(Z,\zeta)
\nonumber \\
\nonumber \\ 
\qquad\qquad\quad &-&\;
\int\limits_X^1 \frac{dZ}{\zeta^2}\; \biggl(1-\frac{\zeta}{2}\biggr)\;
\biggl(\frac{X}{Z} \biggr)
\biggl[
4\biggl(1-\frac{X}{\zeta}\biggr) + \frac{\zeta-2X}{Z}
\biggr]\;\ftil_g(Z,\zeta)
\nonumber \\
\nonumber \\ 
&  & + \: \int_\zeta^1 \frac{dZ}{Z} \;
\frac{(1 -  \zeta/2)(\zeta - X)}{\zeta^2} \; \left [ 
\frac{4 X }{\zeta} +        
\frac{2X - \zeta}{Z} \right ] \; 
\tilde{\cal{F}}_g (Z, \zeta)
\Biggr\}
\nonumber
\eea


\bea
\label{eq:ap6}
\mu\frac{\partial}{\partial \mu}\ftil_g(X,\zeta,\mu)
&=&
\frac{\alpha_S}{\pi} C_F\;
\Biggr\{
\int\limits_0^X dZ\; \biggl(\frac{\Xprime}{\Zprime} \biggr)
\biggl(1-\frac{X}{\zeta} \biggr)\; \frac{\ftil^S(Z,\zeta)}{1-\zeta/2}
+
\int\limits_X^1 dZ \biggl(2-\frac{X^2}{Z \zeta} \biggr)\;
\frac{\ftil^S(Z,\zeta)}{1-\zeta/2}
\nonumber \\
\nonumber \\
& & - \int_\zeta^1 \frac{dZ}{Z}\; 
\frac{(\zeta - X)^2}{\zeta (1 -\zeta/2)} \: \ftil^S(Z, \zeta)
\Biggr\}
\nonumber \\
\nonumber \\
&+&
\frac{\alpha_S}{\pi} N_c\;
\Biggr\{
\int\limits_0^X dZ\;  \biggl(\frac{\Xprime}{\Zprime} \biggr)\;
\biggl[
\frac{2}{\zeta} \biggl(1-\frac{X}{\zeta} \biggr)
\biggl(1+2\frac{X}{\zeta}+\frac{X}{\zeta-Z} \biggr)\; \ftil_g(Z,\zeta) 
\nonumber \\
\nonumber \\
\qquad\qquad\quad &+&\;
\frac{(\Xprime/\Zprime)\; \ftil_g(Z,\zeta)-\ftil_g(X,\zeta)}{X-Z}
\biggr]
\nonumber \\
\nonumber \\ 
&+&
\int\limits_X^1 dZ\;  \biggl(\frac{X}{Z} \biggr)
\biggl[
\frac{2X}{\zeta^2} 
\biggl(3-2\frac{X}{\zeta}+\frac{\zeta-X}{Z} \biggr)\; \ftil_g(Z,\zeta)
\nonumber \\
\nonumber \\ 
&+&
\frac{(X/Z) \ftil_g(Z,\zeta)-\ftil_g(X,\zeta)}{Z-X}
\biggr]
\nonumber \\
\nonumber \\
& + &
\ftil_g(X,\zeta)\;
\biggl[ \frac{11-(2 N_f)/3)}{2N_c}+\ln\frac{X(1-X)}{\zeta}
\biggr]
\\
\nonumber  \\
& +& \int_\zeta^1 \frac{dZ}{Z} 
\frac{(\zeta - X)^2}{Z} \! \left 
[ \frac{1}{X - \zeta + Z} + \frac{2Z}{\zeta^2} 
\left (1 + \frac{2X}{\zeta} +        
\frac{X}{Z} \right ) \right ] \; \ftil_g (Z, \zeta)
\Biggr\}\;.
\nonumber 
\eea
For $\zeta=1$ the above equations reduce to the ERBL evolution equations
for the distribution amplitudes. It is also instructive to check that
both set of equations, (\ref{eq:ap3}) and (\ref{eq:ap6}), 
lead to the same limiting set of equations
when $X \rightarrow \zeta$ from both sides.

The  equations for the singlet $\ftil^S$ and the gluon $\ftil^g$ distributions
form a coupled set of equations which, in general, need to be solved
simultaneously in both the ERBL- and DGLAP-like regions. 
However for  $X>\zeta$ it is sufficient 
to solve the equations only in the DGLAP-like region since the integration
in (\ref{eq:ap3})
involves only parton distributions for values of $Z>X$ (as is true for the
DGLAP equations in the limit $\zeta=0$).
This is not the case if $X<\zeta$. Then the  solutions 
depend on the values of the
parton distributions in the full interval $[0,1]$, and
so both the  set of  equations, (\ref{eq:ap3}) and (\ref{eq:ap6}), 
have to be solved simultaneously.

   
\newpage  
\begin{figure} 
   \vspace*{-1cm} 
    \centerline{ 
     \epsfig{figure=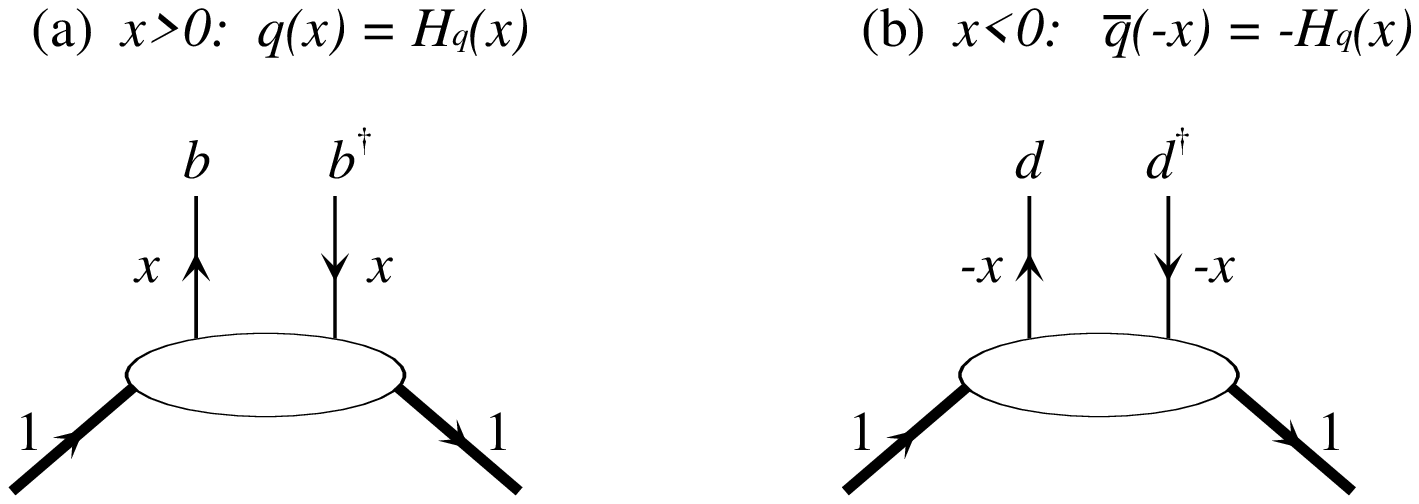,width=15cm} 
               } 
    \vspace*{-0.5cm} 
\caption{Schematic diagrams showing the contributions to $H_q (x)$ with  
$x > 0$ and $x < 0$ respectively, which can be identified with the familiar quark and   
antiquark distributions.  $b, b^\dagger$ are the quark annihilation and creation   
operators and $d, d^\dagger$ are those for the antiquark.  The momentum fractions   
refer to the plus light-cone component of the incoming proton momentum $P$. } 
\label{fig1} 
\end{figure}

\begin{figure} 
   \vspace*{-1cm} 
    \centerline{ 
     \epsfig{figure=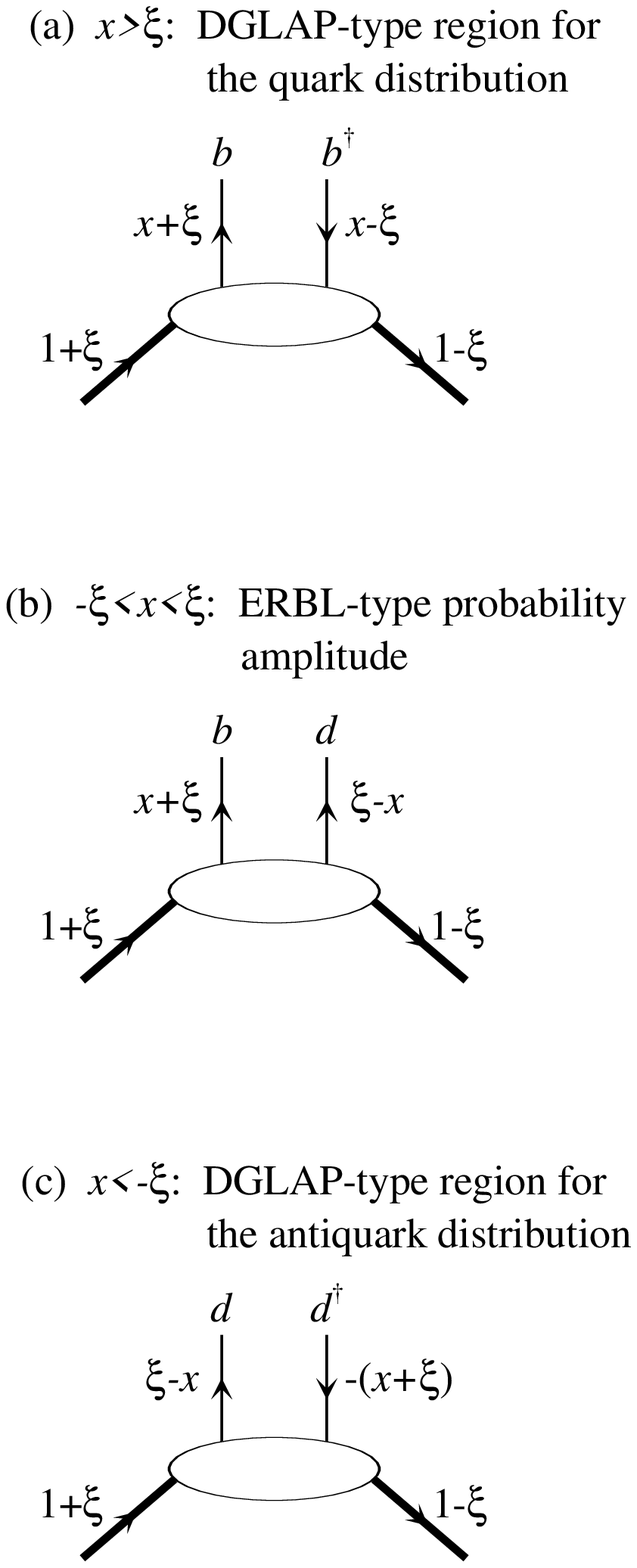,width=13cm} 
               } 
    \vspace*{0.0cm} 
\caption{ 
Schematic diagrams of the off-diagonal distribution  
$H_q (x,\xi)$, in the three distinct  
kinematic regions.  The   
proton and quark momentum fractions refer to $\bar{P}^+$,  
where $\bar{P}$ is the   
average of the incoming and outgoing proton four momentum.   
Note that the four   
momentum transfer satisfies $\Delta^+ = 2 \xi \bar{P}^+$ and  
that $x$ covers the interval $[-1,1]$.} 
\label{fig2} 
\end{figure}  
 
\begin{figure} 
   \vspace*{-1cm} 
    \centerline{ 
     \epsfig{figure=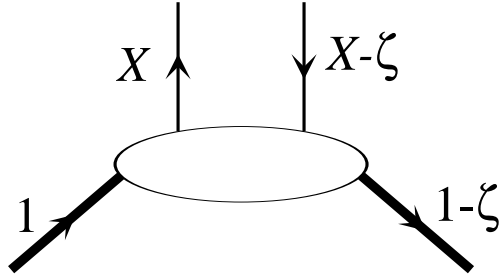,width=8cm} 
               } 
    \vspace*{-0.5cm} 
\caption{The proton and quark momentum fractions with respect to the initial   
proton momentum $P$ corresponding to the off-diagonal distributions             
$\hat{\cal{F}} (X, \zeta)$ defined in the domain $0\le~X\le~1$. 
The four momentum transfer satisfies $\Delta^{+}=\zeta P^{+}$.} 
\label{fig3} 
\end{figure}  
 
\begin{figure} 
   \vspace*{-1cm} 
    \centerline{ 
     \epsfig{figure=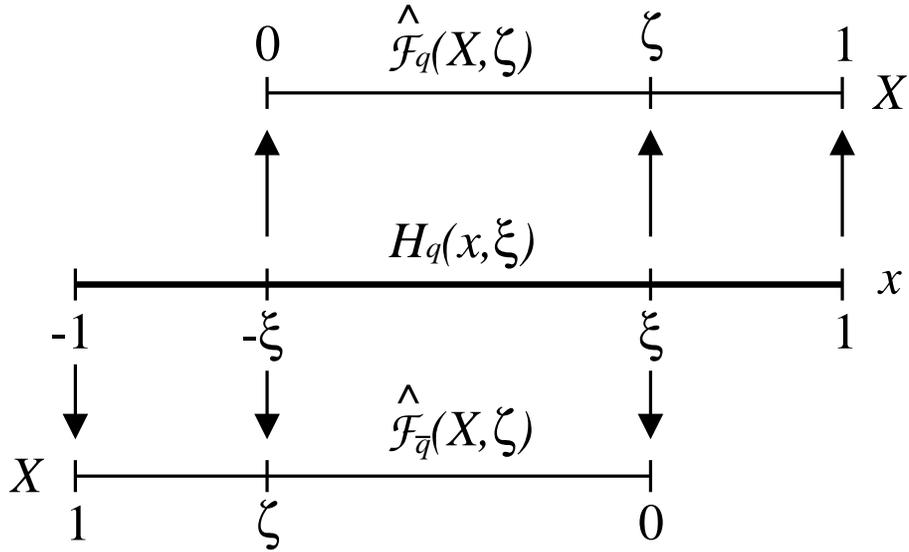,width=13cm} 
               } 
    \vspace*{-0.5cm} 
\caption{A sketch showing how the support $-1 \le x \le 1$ of the off-diagonal   
distribution $H_q$ is translated into the regions $0 \le X \le 1$ of  
the two functions   
$\hat{\cal{F}}_q$ and $\hat{\cal{F}}_{\bar{q}}$.   
The translations are given by   
(\ref{eq:a18}) and (\ref{eq:a19}),  
or by the inverse relations (\ref{eq:a21}).} 
\label{fig4} 
\end{figure}     
      
\begin{figure} 
   \vspace*{-1cm} 
    \centerline{ 
     \epsfig{figure=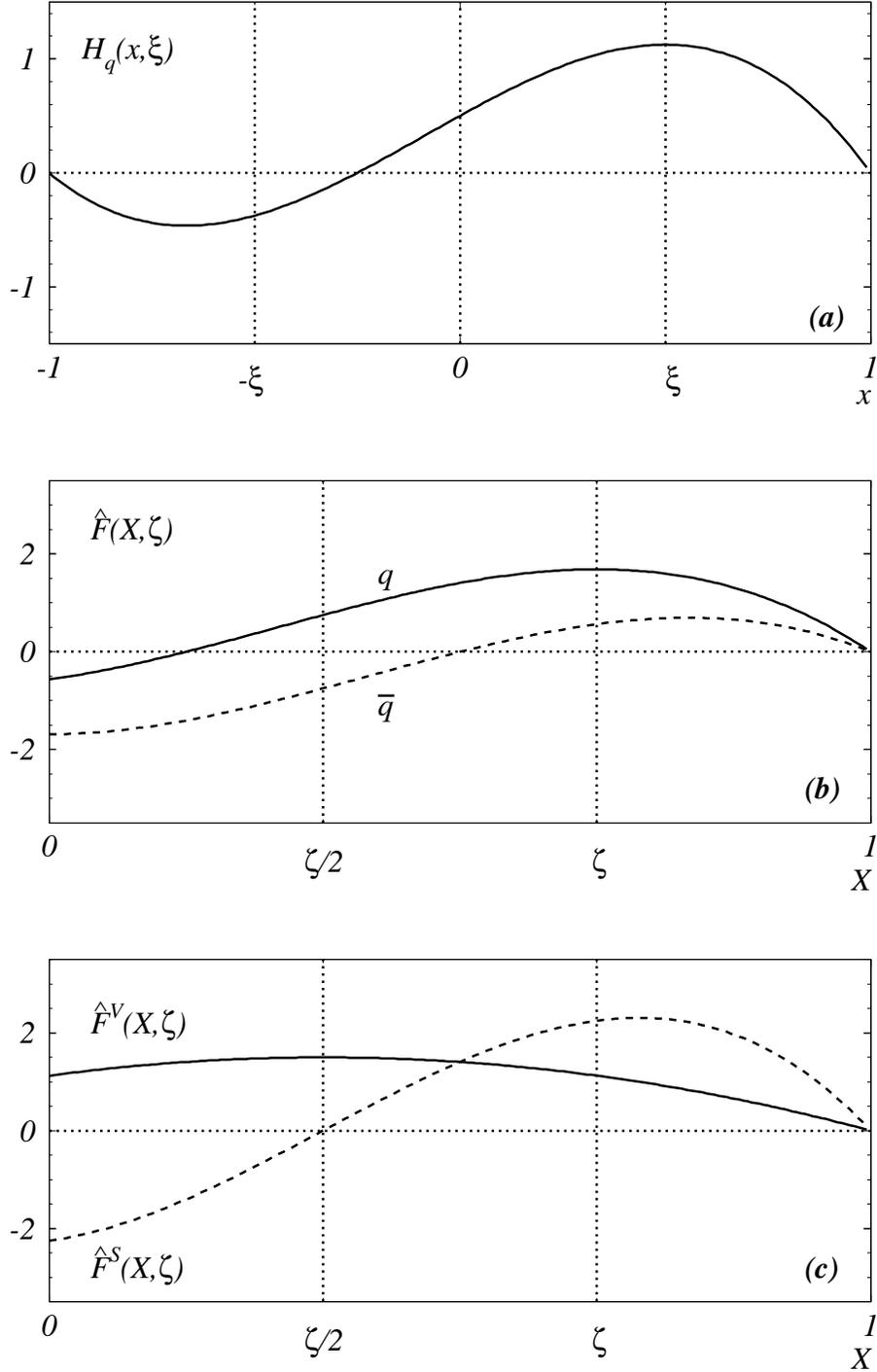,width=15cm} 
               } 
    \vspace*{0.0cm} 
\caption{(a) An example of the  off-diagonal distribution $H_q (x, \xi)$ with $\xi = 
0.5$;    
(b) the distributions $\hat{\cal{F}}_q (X, \zeta)$ and  
$\hat{\cal{F}}_{\bar{q}} (X,\zeta)$ generated from $H_q (x, \xi)$,  
and (c) the resulting non-singlet    
$\hat{\cal{F}}_q^V$ and singlet $\hat{\cal{F}}^S$ distributions showing their    
symmetry and antisymmetry in the ERBL-like region $X < \zeta$.} 
\label{fig5} 
\end{figure}  
 
\begin{figure} 
   \vspace*{-1cm} 
    \centerline{ 
     \epsfig{figure=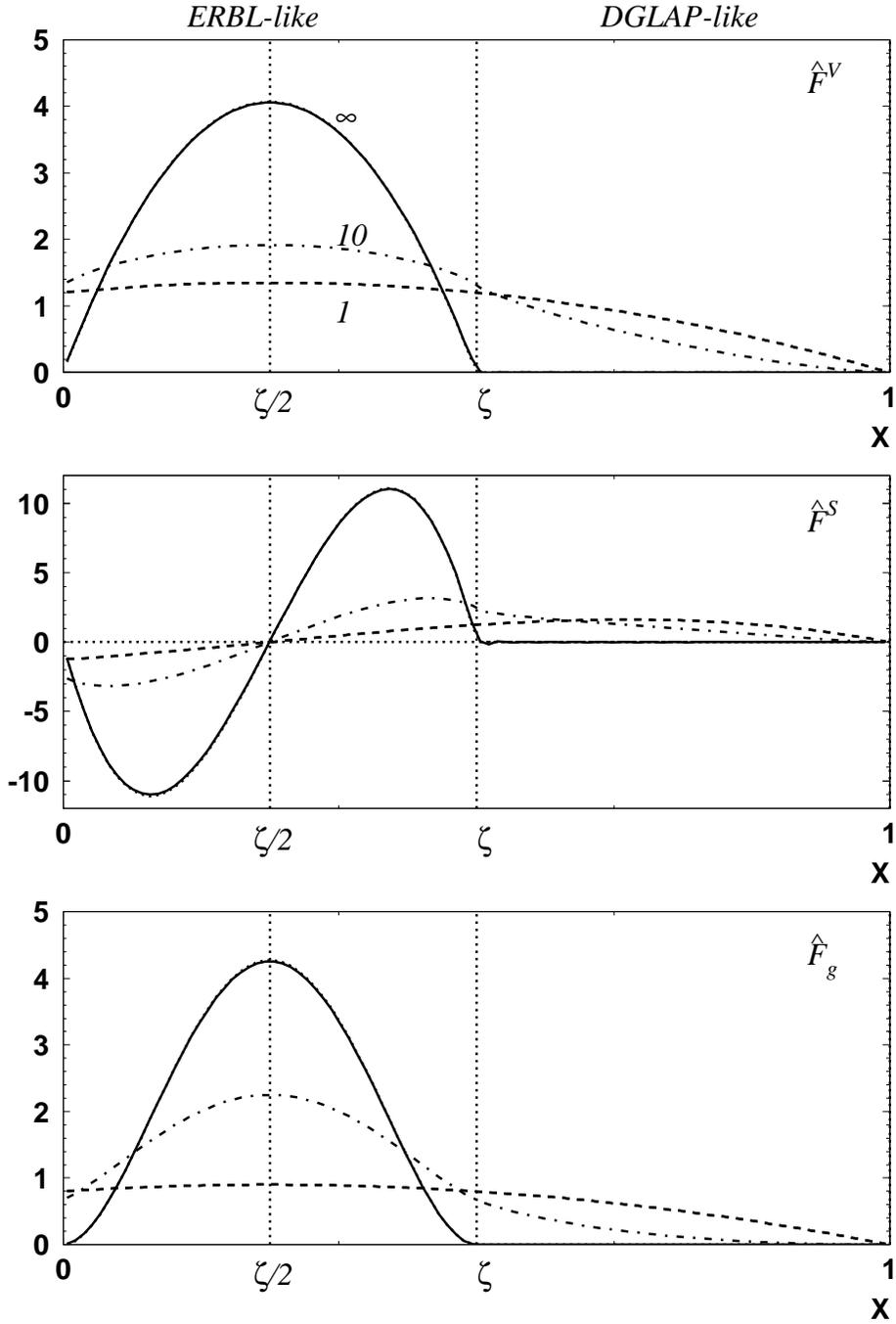,width=15cm} 
               } 
    \vspace*{0.0cm} 
\caption{Evolution of the non-singlet $\hat{\cal{F}}^V_q$, singlet  
$\hat{\cal{F}}^S$ and gluon $\hat{\cal{F}}_g$ distributions defined in the range 
$[0,1]$ from initial input at $\mu=1$~GeV (dashed curves).  
The asymmetry parameter $\zeta=0.5$. The dotted and continuous curves correspond 
to $\mu=10$~GeV and $\mu \rightarrow \infty$ respectively. The latter curves are 
identical to the analytic asymptotic solutions given in (\ref{eq:a33}).}  
\label{fig6} 
\end{figure}

\end{document}